\documentclass[a4paper]{jpconf}
\usepackage{graphicx}

\usepackage{amsfonts,amsthm,amstext,array}
\usepackage{mathrsfs}
\usepackage{bbm}
\usepackage{bm}
\usepackage[arrow,matrix,curve]{xy}

\usepackage{cite}

\usepackage{hyperref}
\usepackage{graphics}
\usepackage{amsmath}
\usepackage{amssymb,amscd}
\usepackage[arrow,matrix,curve]{xy}
\usepackage{bbm}

\usepackage{pdfsync}

\usepackage{epsfig}

\def\={\ =\ }
\def\dd{{\rm d}}

\def\e{{\,\rm e}\,}
\newcommand{\mbf}[1]{{\boldsymbol {#1} }}

\def\ii{{\,{\rm i}\,}}

\newcommand{\bbr}{\mathbb{R}}

\newcommand{\bbc}{\mathbb{C}}

\newcommand{\remark}[1]{}     				
\newcommand{\newsection}{\setcounter{equation}{0}\section}

\begin{document}

\begin{flushright}
\small
EMPG--17--15
\end{flushright}

\title{Magnetic monopoles and nonassociative deformations of quantum theory\footnote{Based on plenary talk
  given at the XXVth International Conference on Integrable Systems
  and Quantum Symmetries, June 6--10, 2017, Prague, Czech Republic; to
be published in {\sl Journal of Physics: Conference Series}.}}

\author{Richard J. Szabo}

\address{Department of Mathematics,
Heriot-Watt University,
Colin Maclaurin Building, Riccarton, Edinburgh EH14 4AS, U.K.}
\address{Maxwell Institute for Mathematical Sciences, Edinburgh, U.K.}
\address{The Higgs Centre for Theoretical Physics, Edinburgh, U.K.}

\ead{R.J.Szabo@hw.ac.uk}

\begin{abstract}
We examine certain nonassociative deformations of quantum mechanics
and gravity in three dimensions related to the dynamics of
electrons in uniform distributions of magnetic charge. We describe a
quantitative framework for nonassociative quantum mechanics in this
setting, which exhibits new
effects compared to ordinary quantum mechanics with sourceless magnetic fields, and the extent to which these theoretical consequences may be
experimentally testable. We relate this theory to noncommutative
Jordanian quantum mechanics, and show that its underlying algebra can
be obtained as a contraction of the alternative algebra of
octonions. The uncontracted octonion algebra conjecturally describes a
nonassociative deformation of three-dimensional quantum gravity
induced by magnetic monopoles, which we propose is realised by a
non-geometric Kaluza-Klein monopole background in M-theory.
\end{abstract}

\newsection{Prologue: Three-dimensional quantum gravity}

Applications of noncommutative geometry in physics are often motivated
as providing a suitable mathematical framework for describing the modifications of spacetime geometry
at very short distance scales which are expected in a
quantum theory of gravity; the length scale at which such effects
become important is usually understood to be the Planck length $\ell_{\rm
  P}$. However, there are relatively few precise quantitative
connections between noncommutative geometry and models of quantum
gravity. One notable exception is quantum gravity in three spacetime
dimensions; that three-dimensional quantum gravity naturally implies a
``fuzziness'' to short distance spacetime structure
was noted already in the early work of 't~Hooft~\cite{tHooft:1996ziz}
and others, see e.g.~\cite{Matschull:1997du}, who observed that the
spectrum of position operators in this theory is discrete. In this paper we shall be
primarily concerned with the precise realisation of these structures
obtained by~\cite{Freidel:2005me}, who consider a Ponzano-Regge spin foam
model of three-dimensional quantum gravity coupled to spinless matter
fields. After integrating out the gravitational degrees of freedom in
this model, they obtain an effective scalar field theory on a
noncommutative spacetime described via the deformed phase space
commutation relations
\begin{eqnarray}
\big[x^i\,,\,x^j\big]&=&\ii{\ell_{\rm P}}\,\varepsilon^{ijk}\,
x^k  \ , \nonumber \\[4pt]
\big[x_i\,,\,p_j\big]&=&\ii\sqrt{\hbar^2-{\ell_{\rm P}^2}
\,p^2}~\delta_{ij}-
\ii{\ell_{\rm P}}\,\varepsilon_{ijk}\,p_k \ , 
                             \nonumber \\[4pt]
\big[p_i\,,\,p_j\big]&=&0 \ .
\label{eq:3DQGrels}\end{eqnarray}
The commutation relations among position coordinates $x^i$ specify a
Lie algebra noncommutative spacetime; the relations apply to both
Euclidean signature, wherein the pertinent Lie group is $SU(2)$ and
which is the main focus of this paper, and also in Minkowski signature
wherein the pertinent Lie group is $SO(1,2)$.

The commutation relations \eqref{eq:3DQGrels} can be understood in the
following way. A peculiar feature of quantum gravity in three
dimensions is that the momenta $p_i$ of a particle are bounded from
above, with the bound set by the inverse of the Planck length
$\ell_{\rm P}$. In particular, we can take the momentum space to be a
sphere of radius $\ell_{\rm P}^{-1}$. Then the commutators in
\eqref{eq:3DQGrels} simply reflect the fact that position coordinates
act as derivations on this sphere, as implied by the usual canonical
commutation relations. The deformation of the latter in
\eqref{eq:3DQGrels} makes the noncommutative algebra invariant under a
$\kappa$-deformation of the Poincar\'e group in three
dimensions, with $\kappa=\ell_{\rm P}^{-1}$, such that the phase space commutation relations define a
noncommutative but associative algebra. Setting $\ell_{\rm P}=0$ corresponds to turning off the effects of quantum
gravity and leaves an undeformed canonical phase space in
\eqref{eq:3DQGrels} .

Many interesting physical properties of three-dimensional quantum
gravity can be infered from the relations \eqref{eq:3DQGrels}. For
example, they imply a deformation of the usual addition law for
Fourier wavenumbers, which in Minkowski signature leads to the modified
dispersion relations
\begin{eqnarray}
E^2\=\mbf p^2\,c^2-\Big(\frac{\sinh({\ell_{\rm P}\,\hbar^{-1}}\,m\,c^2)}
{{\ell_{\rm P}\,\hbar^{-1}}}\Big)^2 \ .
\end{eqnarray}
In the limit $\ell_{\rm P}\to0$ wherein quantum gravity effects are
neglected, this is just the usual relativistic dispersion law for a
scalar particle in three spacetime dimensions. The deformation by $\ell_{\rm
  P}\neq0$ is interpreted as the statement that doubly special relativity arises precisely
in the low energy limit of three-dimensional quantum
gravity~\cite{Freidel:2005me}.

The purpose of this paper is to describe a novel conjectural
nonassociative deformation of the three-dimensional quantum gravity
algebra \eqref{eq:3DQGrels} which is implied by certain magnetic dual
analogues of the types of nonassociative spacetime geometries that are anticipated to arise in
non-geometric string theory and M-theory, see
e.g.~\cite{Blumenhagen:2010hj,Lust:2010iy,Blumenhagen:2011ph,Mylonas:2012pg,Bakas:2013jwa,Blumenhagen:2013zpa,Aschieri:2015roa,Gunaydin:2016axc,Kupriyanov:2017oob,Lust:2017bgx}. For
a while we shall therefore leave the present setting of
three-dimensional quantum gravity and explain how nonassociative
deformations of phase space naturally appear in a simpler elementary quantum
mechanical setting of magnetic monopole distributions (Section~\ref{sec:MagneticQM}). By attempting to reconcile this
particular nonassocative deformation of quantum mechanics with the
nonassocative algebras of observables proposed in the early days of
quantum mechanics by Jordan, von~Neumann, Wigner and others to study
the mathematical and conceptual foundations of quantum theory, we
shall arrive at a new framework that conjecturally describes magnetic
monopoles in the spacetime of three-dimensional quantum gravity
(Section~\ref{sec:Jordan}). We conclude with a proposed realisation of
such a system within the framework of M-theory (Section~\ref{sec:KKmonopole}).

\newsection{Magnetic monopoles and nonassociative quantum
  mechanics\label{sec:MagneticQM}}

We shall start by discussing the quantum mechanics of 
electrons propagating in sources of magnetic charge, focusing on the
case of uniform monopole distributions which are the suitable analog
systems dual to certain non-geometric string backgrounds. In this
latter case we argue that the conventional framework of quantum theory
nessitates a nonassociative deformation, and then demonstrate that
such a model of nonassociative quantum mechanics is not only possible,
but physically sensible
and potentially testable.

\subsection{Magnetic sources in quantum mechanics}

Electrons propagating in a magnetic field $\mbf B(\mbf x)$ under the influence of an external potential $V(\mbf x)$ in
  three dimensions
  classically obey the Lorentz force law
\begin{eqnarray}
\dot{\mbf  p}\= \frac e{2\, m\,c} \, \big({\mbf
     p}\times {\mbf B} -\mbf B\times\mbf p\big) +\mbf\nabla V(\mbf x)
\label{eq:Lorentzforce}\end{eqnarray}
where $\mbf p=m\,\dot{\mbf x}$ is the kinematical
  momentum (as opposed to the canonical momentum), and we have written
  the right-hand side in a form that generalises to the quantum theory. The corresponding Hamiltonian on
  phase space is the sum of kinetic and potential energies:
\begin{eqnarray}
\mathsf{H}(\mbf x,\mbf p) \= \frac1{2m}\, {\mbf  p}^2+V(\mbf x) \ .
\label{eq:Hamgeneral}\end{eqnarray}
Then the quantum mechanical Heisenberg equations of motion
\begin{eqnarray}
-\ii\hbar \, \dot{\mbf p}\=
  [\mathsf{H},{\mbf p}] \quad , \quad -\ii\hbar\, \dot{\mbf x} \=
  [\mathsf{H}, {\mbf x}]
\end{eqnarray}
are compatible with the Lorentz force law and the relation $\dot{\mbf
  x}={\mbf p}/m$ only with the phase
space commutation relations
\begin{eqnarray}
[x^i,x^j]\=0 \quad \ , \quad [x^i, p_j]\=\ii\hbar\, \delta^i{}_j
\quad , \quad [ p_i, p_j]\= \frac{\ii \hbar\, e}c \,
\varepsilon_{ijk}\, B^k \ ,
\label{eq:Lorentzcommrels}\end{eqnarray}
which deform the canonical commutators to a noncommutative momentum
space. Note that here and below everything is formulated without any reference to a vector potential, so that these considerations apply as well to the cases where the magnetic field $\mbf B$ has sources.

Globally, the noncommutativity can be described in terms of the
magnetic translation operators
\begin{eqnarray}
U({\mbf a})\=\exp\big(\mbox{$\frac\ii\hbar$} \, {\mbf a}\cdot{\mbf
     p} \big)
\end{eqnarray}
which generate finite translations $U(\mbf a)\,\mbf x\, U(\mbf a)^{-1}=\mbf
x+\mbf a$ due
to the second commutator in \eqref{eq:Lorentzcommrels}. A simple calculation shows that they
 do not commute~\cite{Zak:1964zz}:
\begin{eqnarray}
U({\mbf a}_1)\, U({\mbf a}_2)\= \e^{\frac{\ii e}{\hbar\, c}\,
  \Phi_{2}(\mbf x;\mbf a_1,\mbf a_2)}\, U({\mbf
  a}_1+{\mbf a}_2)\= \e^{\frac{2\ii e}{\hbar\,c}\, \Phi_{2}(\mbf
  x;\mbf a_1,\mbf a_2)}\,U({\mbf a}_2)\, U({\mbf a}_1) \ ,
\label{eq:Uacommrels}\end{eqnarray}
where
\begin{eqnarray}
\Phi_2(\mbf x;\mbf a_1,\mbf a_2)\=\int_{\langle \mbf a_1,\mbf
  a_2\rangle_{\mbf x}}\, \mbf B\cdot\dd\mbf S
\end{eqnarray}
is the magnetic flux
  through the plane of the triangle {$\langle{\mbf a}_1,{\mbf
  a}_2\rangle_{\mbf x}$} based at $\mbf x\in\bbr^3$ and spanned by
the vectors $\mbf a_1$ and $\mbf a_2$. 

One can define a 3-bracket on momentum space as the combination of
iterated commutators
\begin{eqnarray}
[ p_1,  p_2, p_3 ] \ := \ [ p_1,[ p_2, p_3]] +
[ p_2,[ p_3, p_1]] + [ p_3,[ p_1, p_2]] 
\end{eqnarray}
that would vanish if the brackets \eqref{eq:Lorentzcommrels} satisfied
the Jacobi identity; it is called the `Jacobiator'. One easily computes in this case
\begin{eqnarray}
[ p_1,  p_2, p_3 ] \= \frac {\hbar^2\,e}c \,
\mbf\nabla\cdot{\mbf B} \ ,
\label{eq:mom3bracket}\end{eqnarray}
which expresses a further deformation of canonical phase space to a
nonassociative momentum space. Globally, one easily computes using
\eqref{eq:Uacommrels} that the magnetic translation operators thus do not associate:
\begin{eqnarray}
\big(U({\mbf a}_1)\, U({\mbf a}_2) \big)\, U({\mbf a}_3 )\= \e^{\frac{\ii
  e}{\hbar\,c} \, \Phi_{3}(\mbf x;\mbf a_1,\mbf a_2,\mbf a_3)}\, U({\mbf a}_1)\, \big(U({\mbf a}_2) \, U({\mbf
  a}_3) \big) \ ,
\label{eq:Uassocrels}\end{eqnarray}
where $\Phi_{3}(\mbf x; \mbf a_1,\mbf a_2,\mbf a_3)$ is the sum of the
magnetic fluxes through $\langle \mbf a_2,\mbf a_3\rangle_{\mbf x+\mbf
  a_1}$ and $\langle \mbf a_1,\mbf a_2+\mbf a_3\rangle_{\mbf x}$,
minus the fluxes through $\langle\mbf a_1,\mbf a_2\rangle_{\mbf x}$
and $\langle\mbf a_1+\mbf a_2,\mbf a_3\rangle_{\mbf x}$. Together this
gives the magnetic flux through the faces of the tetrahedron {$\langle {\mbf a}_1,{\mbf
  a}_2, {\mbf a}_3\rangle_{\mbf x}$} based at $\mbf x$ and spanned by
the vectors $\mbf a_1$, $\mbf a_2$ and $\mbf a_3$, which using the
divergence theorem can be written as the magnetic charge
\begin{eqnarray}
\Phi_{3}(\mbf x; \mbf a_1,\mbf a_2,\mbf a_3)\=\int_{\langle \mbf a_1,\mbf
  a_2,\mbf a_3\rangle_{\mbf x}}\, \mbf\nabla\cdot\mbf B \ \dd V
\end{eqnarray}
enclosed by the tetrahedron. There are three
generic situations that can now arise.

First, in Maxwell theory without
magnetic sources, we have $\mbf\nabla\cdot\mbf B=0$. In this case all variables associate. The commutation relations
\eqref{eq:Lorentzcommrels} may then be represented on the Hilbert
space $L^2(\bbr^3)$ of wavefunctions $\psi(\mbf x)$ in a
Schr\"odinger polarization by the kinematical momentum operators ${\mbf p}=-\ii\hbar\, \mbf\nabla-\frac ec\, {\mbf
      A}$, where $\mbf A$ is a globally defined vector potential for the magnetic
    field: $\mbf B=\mbf\nabla\times\mbf A$. The vanishing of the
    magnetic charge, $\Phi_{3}=0$, implies that the phase factor
    $\Phi_2$ defines a 2-cocycle of the translation group of
    $\bbr^3$, and the
relations \eqref{eq:Uacommrels} reflect the fact that the
magnetic translation operators only form a projective representation
of this group; as usual, such projective actions on
wavefunctions $\psi(\mbf x)$ are well-defined in quantum mechanics.

Next, suppose that there are magnetic sources so that
$\mbf\nabla\cdot\mbf B\neq0$. Then the magnetic translation
operators generate a nonassociative algebra unless the magnetic charge
obeys
\begin{eqnarray}
\frac{e\,\Phi_{3}}{2\pi\,\hbar\,c}\= \mbox{integer} \ .
\label{eq:Diracquant}\end{eqnarray}
In this case $\mbf\nabla\cdot\mbf B$ is necessarily a sum of
delta-functions, representing a set of isolated point-like magnetic
monopoles, or else the quantization condition \eqref{eq:Diracquant}
would be incompatible with continuous variations of the position vectors $\mbf
a_i\in\bbr^3$. This constraint is simply the celebrated Dirac
quantization condition~\cite{Dirac}, as first pointed out in the present context by
Jackiw~\cite{Jackiw:1984rd}; with it, the magnetic translations may be
represented by linear operators on a Hilbert space, which necessarily
associate. At the infinitesimal level, the Jacobiator
\eqref{eq:mom3bracket} is only non-vanishing at the loci of the
magnetic sources. For isolated magnetic monopoles, one can excise
their locations from position space and quantize the system in the
standard way on the excised space: Classically the electrons
never reach the monopole locations because of angular
momentum conservation~\cite{Bakas:2013jwa}, while at the quantum level the wavefunctions vanish
at the magnetic charge loci. Moreover, the corresponding local
vector
potentials $\mbf A$, and the magnetic field $\mbf
B\=\mbf\nabla\times\mbf A$, are singular at the monopole locations
which can be interpreted in the standard way in terms of the
singularities of Dirac strings which emanate from the
monopoles~\cite{Dirac}; for example, a single Dirac monopole sources the magnetic field
\begin{eqnarray}
\mbf B(\mbf x)\= \frac{\Phi_3}{4\pi} \,\frac{\mbf x}{|\mbf x|^3} \ .
\end{eqnarray}
Geometrically, the wavefunctions $\psi(\mbf x)$ can be treated as sections and the
local monopole
potential $\mbf A$ as a connection of a non-trivial
$U(1)$-bundle over the excised space~\cite{WuYang}, whose degree is precisely the
magnetic charge \eqref{eq:Diracquant}.

In this paper we are interested in the case of a homogeneous magnetic
charge density, wherein $\mbf\nabla\cdot\mbf B$ is constant. In this
case one cannot excise the support of the magnetic charge
distribution, or else one would be left with empty space, and one
needs to deal directly with the nonassociativity of momentum
coordinates and the magnetic translations. This is a
manifestation of the type of `local non-geometry' that arises in flux
compactifications of string theory; it originates from the fact
that the uniform distribution of magnetic charge can no longer be
described by a connection on a $U(1)$-bundle over a non-contractible
space, as now a vector potential $\mbf A$ does not even exist locally
when $\mbf\nabla\cdot\mbf B\neq0$ everywhere, but rather it induces a
connection $F_{ij}=\varepsilon_{ijk}\, B^k$ on a (trivial) $U(1)$-gerbe on
$\bbr^3$. In particular, there is no relation between the kinematical
and canonical momentum in this case. Moreover, the magnetic flux $\Phi_2$ is a
2-cochain which no longer defines a (weak) projective representation; its coboundary is the magnetic charge $\Phi_3$
which defines a 3-cocycle of the translation group. Nonassociativity that is induced by a 3-cocycle controlling the
associativity of operators can be elegantly dealt with by working in a suitable braided monoidal category of representations~\cite{Barnes:2014ksa}. The geometric meaning of
such higher structures in quantization is described in~\cite{Bunk:2016rta,Bunk:2016gus}, where the
sections of the trivial $U(1)$-gerbe on $\bbr^3$ with connection given by
the magnetic field $\mbf B(\mbf x)=\frac13\,\mbf x$ are shown to correspond to
a certain 2-Hilbert space of Hermitian matrix-valued one-forms on
$\bbr^3$ with bicovariantly constant matrix-valued functions on $\bbr^3$ as morphisms.

Previous algebraic characterisations of the commutation
relations \eqref{eq:Lorentzcommrels} have discussed their realisation
as a nonassociative Malcev
algebra~\cite{Gunaydin:1985ur,Gunaydin:2013nqa}, and 
more generally as an alternative (or Jordan)
algebra~\cite{Bojowald:2014oea}; we shall come back to these points in
Section~\ref{sec:Jordan}, and indicate why these
realisations are somewhat subtle from the perspective of this paper. Classically,
the Lorentz force equations \eqref{eq:Lorentzforce} no longer seem to be integrable, as the
conservation of angular momentum of the Dirac monopole background is
lost~\cite{Bakas:2013jwa}, while at the quantum level it is clear that
some formalism of \emph{nonassociative quantum mechanics} is needed; in the
rest of this section we describe such a framework.

\subsection{Phase space quantum mechanics}

Operators which act on a separable Hilbert space necessarily
associate, by definition. Hence the
standard operator-state techniques of quantum mechanics are inadequate
to handle systems with a
non-trivial 3-cocycle that obstructs associativity. This can be
presumably dealt with by a suitable higher operator-state formalism
adapted to the quantum 2-Hilbert space of~\cite{Bunk:2016gus}, but
such techniques have not yet been developed. We can,
however, appeal to the phase space formulation of quantum mechanics,
which was succinctly developed by Gr\"onewold~\cite{Gronewold},
Moyal~\cite{Moyal} and others, following earlier work of
Weyl~\cite{Weyl} and Wigner~\cite{Wigner}; see
e.g.~\cite{Zachos:2001ux} for a pedagogical introduction. One of the premises of this
formalism is to treat position and momentum on equal footing, which
are usually otherwise treated
asymmetrically when choosing e.g.~a Schr\"odinger polarization. It is through this 
that the formalism of star products was originally introduced.

Let us recall the basic dictionary in the case where {$\mbf B=\mbf
  0$}.  Operators in general become functions on phase space, while
observables correspond to real
functions. Traces of operators are given by integrating functions over
phase space. States {$\psi$} become real phase space quasi-probability
distribution (Wigner) functions which are defined by Fourier
transformation of their position space representations as
\begin{eqnarray}
W_\psi(\mbf x,\mbf p)\=  \int\, \frac{\dd \mbf y}{(2\pi)^3} \ \big\langle
  \mbf x+\mbox{$\frac\hbar2$}\, \mbf
  y\,\big|\,\psi\big\rangle\big\langle\psi\,\big|\,\mbf
  x-\mbox{$\frac\hbar2$}\,\mbf y \big\rangle \, \e^{-\ii \mbf
  y\cdot\mbf p} \ ,
\end{eqnarray}
such that the expectation value of an operator (function) is computed as
\begin{eqnarray}
\langle A\, \rangle_\psi\= \int\, \dd \mbf x \ \dd \mbf p \
  W_\psi(\mbf x,\mbf p)\, A(\mbf x,\mbf p) \ .
\end{eqnarray}
The operator product, which captures the crucial noncommutativity of quantum
mechanics that leads to e.g. uncertainty relations, becomes the associative noncommutative Moyal star product
\begin{eqnarray}
A(\mbf x,\mbf p)\star B(\mbf x,\mbf p)\=A\big(\mbf
  x-\mbox{$\frac{\ii\hbar}2\,\mbf\nabla_{\mbf p}$}\,,\, \mbf
  p+\mbox{$\frac{\ii\hbar}2\,\mbf\nabla_{\mbf x}$}\big)B(\mbf x,\mbf p)
\label{eq:Moyalstar}\end{eqnarray}
which is defined by replacing the arguments of a function
$A(\mbf x,\mbf p)$ by their Bopp shifts and letting the resulting
differential operator act on the function $B(\mbf x,\mbf p)$; this can
be properly defined when $A(\mbf x,\mbf p)$ is a polynomial and then
continuing to power series expansions. 

Given a Hamiltonian
$\mathsf{H}(\mbf x,\mbf p)$ on phase space, dynamics of observables is
governed by the Heisenberg-type time evolution equations
\begin{eqnarray}
 \frac{\dd A}{\dd t} \= \frac{ [A,{\sf H}]_{\star}}{\ii\hbar}
\end{eqnarray}
where the commutator is computed using the star product
\eqref{eq:Moyalstar}, $[A,B]_\star = A\star B-B\star A$.
Stationary states change simply by a time-dependent phase and
are defined via the
`star-genvalue equation'
\begin{eqnarray}
{\sf H}\star W_\psi\=E\, W_\psi \ ,
\label{eq:stargenvalue}\end{eqnarray}
which is the time-independent Schr\"odinger equation determining the
energy eigenvalues $E$.
This approach to quantization can successfully capture the quantum
mechanics of the standard textbook examples, but it also has various
conceptual subtleties and limitations; for example, Wigner functions in general
need not be positive and so do not genuinely determine probability
distributions.

\subsection{Nonassociative quantum mechanics}

Following~\cite{Mylonas:2013jha} we can apply the phase space formulation to develop a version of
nonassociative quantum mechanics that rather remarkably passes all
basic physical consistency tests, and moreover has great quantitative
power to compute new physical consequences of nonassociativity. Developing
nonassociative quantum mechanics also brings into question many foundational
issues and can teach us a lot about the nature of quantum theory itself. We
shall work throughout with the case where $ \rho := \frac ec\,
\mbf\nabla\cdot\mbf B$ is a constant monopole density, and choose the
gauge 
\begin{eqnarray}
\mbf B(\mbf x) \= \frac{\rho\,c}{3\,e}\, \mbf x \ .
\end{eqnarray}

The first ingredient we need is a suitable modification of the
Moyal star product \eqref{eq:Moyalstar}. Such a nonassociative star
product was first constructed in~\cite{Mylonas:2012pg}. Here we shall
use the form developed in~\cite{Mylonas:2013jha} (see also~\cite{Kupriyanov:2015dda}), and hence introduce
the nonassociative phase space ``monopole star product'' via a
simple modification of the Bopp shifts in \eqref{eq:Moyalstar} to 
`$\rho$-twisted Bopp shifts' as
\begin{eqnarray}
A(\mbf x,\mbf p)\star B(\mbf x,\mbf p)\=A\big(\mbf
  x-\mbox{$\frac{\ii\hbar}2\,\mbf\nabla_{\mbf p}$}\,,\, \mbf
  p+\mbox{$\frac{\ii\hbar}2\,(\mbf\nabla_{\mbf x}+ \frac13\,\rho\, \mbf
  x\times \mbf\nabla_{\mbf p})$}\big)B(\mbf x,\mbf p) \ .
\label{eq:monopolestar}\end{eqnarray}
That this provides a suitable quantization of the magnetic monopole
algebra \eqref{eq:Lorentzcommrels} and \eqref{eq:mom3bracket} can be
seen by computing the corresponding star commutators of phase space
coordinate functions
\begin{eqnarray}
[x^i,x^j]_\star\=0 \quad \ , \quad [x^i, p_j]_\star\=\ii\hbar\, \delta^i{}_j
\quad , \quad [ p_i, p_j]_\star\= \frac{\ii \hbar}3 \,  \rho\,
  \varepsilon_{ijk}\, x^k 
\end{eqnarray}
and the corresponding non-vanishing star Jacobiator
\begin{eqnarray}
[p_1,p_2,p_3]_\star\=\hbar^2\,  \rho \ .
\label{eq:p123star}\end{eqnarray}
Globally, the monopole star product reproduces the appropriate algebra \eqref{eq:Uacommrels} and \eqref{eq:Uassocrels} of magnetic translation operators as
\begin{eqnarray}
U({\mbf a}_1)\star U({\mbf a}_2)\= \e^{\frac{\ii}{\hbar}\,
  \Phi_{2}(\mbf x;\mbf a_1,\mbf a_2)}\, U({\mbf
  a}_1+{\mbf a}_2) \quad , \quad \Phi_{2}(\mbf x;\mbf a_1,\mbf a_2)\= \frac16\, \rho\, (\mbf a_1\times\mbf a_2)\cdot\mbf x
\end{eqnarray}
and
\begin{eqnarray}
\big(U({\mbf a}_1)\star U({\mbf a}_2) \big)\star U({\mbf a}_3 ) &=& \e^{\frac{\ii
 }{\hbar} \, \Phi_{3}(\mbf x;\mbf a_1,\mbf a_2,\mbf a_3)}\, U({\mbf a}_1)\star \big(U({\mbf a}_2) \star U({\mbf
  a}_3) \big) \quad , \nonumber \\ && \qquad \qquad\Phi_{3}(\mbf x;\mbf a_1,\mbf a_2,\mbf a_3) \= \frac{\hbar^3}6\, \rho \, \mbf a_1\cdot(\mbf a_1\times\mbf a_2) \ .
\end{eqnarray}
The formula \eqref{eq:monopolestar} is exact when $\rho$ is constant.

The monopole star product has various noteworthy properties which are
important for calculations in nonassociative quantum mechanics. First
of all, since the star product 
$ A\star B$ differs from the ordinary pointwise product of functions
$A\, B$ by total derivative terms, it is `2-cyclic' in the sense that
\begin{eqnarray}
\int\, \dd\mbf x \ \dd\mbf p \  A\star B\=\int\, \dd\mbf x \ \dd\mbf p
  \ B\star A\= \int\, \dd\mbf x \ \dd\mbf p \  A\,B \ ,
\label{eq:2cyclic}\end{eqnarray}
for suitable functions $A$ and $B$ of Schwartz class. Hence noncommutativity at this order is washed away upon
integration, i.e. ``on-shell''. Similarly, since the triple star product
$ A\star (B\star C)$ differs from $(A\star B)\star C$ by total
derivative terms, it is `3-cyclic' in the sense that 
\begin{eqnarray}
\int\, \dd\mbf x \ \dd\mbf p \  A\star (B\star
    C)\=\int\, \dd\mbf x \ \dd\mbf p \ (A\star B)\star C \ . 
\end{eqnarray}
Hence nonassociativity at this order is also absent on-shell. However,
this is not true for higher order multiple star products in general,
see~\cite{Mylonas:2013jha} for a general analysis of this
feature. Finally, the monopole star product is Hermitian in the sense
that it mimicks the usual conjugation properties of operator products,
\begin{eqnarray}
(A\star B)^* \= B^*\star A^* \ , 
\end{eqnarray}
and it is unital in the sense that the
constant function $1$ serves as an identity element for the star
product algebra of functions,
\begin{eqnarray}
A\star 1\=A\=1\star A \ .
\label{eq:unital}\end{eqnarray}

A {state} in nonassociative quantum
mechanics is specified by a collection of $L^2$-normalized phase space
wavefunctions {$\psi_a(\mbf x,\mbf p)$},
\begin{eqnarray}
\int\, \dd\mbf x \ \dd\mbf p \ \big|\psi_a(\mbf x,\mbf p)\big|^2\=1 \ , 
\end{eqnarray}
together with statistical probabilities {$\mu_a\in[0,1]$},
\begin{eqnarray}
\sum_a\,\mu_a\=1 \ . 
\end{eqnarray}
Expectation values are then defined by 
\begin{eqnarray}
\langle A\rangle_\psi\= \sum_a \, \mu_a \ \int\, \dd\mbf x \ \dd\mbf p \ 
\psi_a^* \star (A\star \psi_a) \ ,
\label{eq:NAexpvalue}\end{eqnarray}
which using 2-cyclicity and 3-cyclicity of the monopole star product
can be written as
\begin{eqnarray}
\langle A\rangle_\psi \= \int\, \dd\mbf x \ \dd\mbf p \  W_\psi(\mbf x,\mbf p)\, A(\mbf x,\mbf p)
\end{eqnarray}
where 
\begin{eqnarray}
W_\psi\=\sum_a\, \mu_a\, \psi_a\star\psi_a^* \quad , \quad \int\,
  \dd\mbf x \ \dd\mbf p \ W_\psi(\mbf x,\mbf p)\=1
\end{eqnarray}
plays the role of
a Wigner distribution function and should be thought of here as a `density
operator'.

Now let us demonstrate that some of the basic physical requirements of
quantum theory
are satisfied by this setup. Let us first check reality of measurements. With the
definition \eqref{eq:NAexpvalue}, using Hermiticity and 3-cyclicity we
have
\begin{eqnarray}
\langle A \rangle_\psi^* \= \sum_{a} \, \mu_a \ \int\, \dd\mbf x \ \dd\mbf p \  (A \star \psi_a)^* \star \psi_a
\= \sum_{a} \, \mu_a \ \int\, \dd\mbf x \ \dd\mbf p \  \psi_a^* \star (A^* \star \psi_a) \= \langle A^* \rangle_\psi
\end{eqnarray}
and hence observables $A=A^*$ have real expectation values; thus
physical measurements are real in this quantum theory. A similar
calculation establishes positivity of measurements~\cite{Mylonas:2013jha}.

Next we show that observables {$A=A^*$} have
real eigenvalues. Using Hermiticity the conjugate of the star-genvalue
equation {$A\star W_\psi=\lambda\, W_\psi$} is $W_\psi^* \star
A=\lambda^*\, W_\psi^*$ which gives
\begin{eqnarray}
W_\psi^* \star (A \star W_\psi) - (W_\psi^* \star A) \star W_\psi \= (\lambda -
\lambda^*)\, (W_\psi^* \star W_\psi) \ .
\end{eqnarray}
In the associative case we would be done at this stage, because the
left-hand side of this equation would vanish, while the right-hand
would be generically non-zero, but this is not so in the
nonassociative case. However, the left-hand side vanishes on-shell by
3-cyclicity, so by integrating both sides we get
\begin{eqnarray}
0 \= (\lambda-\lambda^*) \, \int\, \dd\mbf x \ \dd\mbf p \  W_\psi^* \star W_\psi \= (\lambda-\lambda^*) \, \int\, \dd\mbf x \ \dd\mbf p \ \big|W_\psi(\mbf x,\mbf p) \big|^2
\end{eqnarray}
and now the right-hand side is non-zero unless $\lambda=\lambda^*$. These
calculations illustrate the general features of checks in
nonassociative quantum mechanics: In all cases the calculations
proceed as in the associative case, but there are always a few extra
steps required due to nonassociativity. But against all odds, the
properties \eqref{eq:2cyclic}--\eqref{eq:unital} of the monopole star
product ensure that this
nonassociative deformation of quantum theory passes all
consistency checks. Various other basic features, including quantum
uncertainty relations, are worked out in~\cite{Mylonas:2013jha}.

The key feature that makes such a nonassociative deformation tractable is the occurence of a 3-cocycle controlling associativity, as mentioned before; such an algebra should be more properly refered to as `quasi-associative', since it would be hopeless to try to make things work in an arbitrary nonassociative algebra. In the present case it means that there is a multiplicative associator $\Phi$ which controls the rebracketing of triples of phase space functions under the monopole star product \eqref{eq:monopolestar} as~\cite{Mylonas:2013jha,Aschieri:2015roa}
\begin{eqnarray}
{A}\star( {B}\star
 {C}) \ \xrightarrow{ \
  \Phi_{{A},{B}, {C}} \ } \
({A}\star {B})\star {C} \ , 
\end{eqnarray}
which is most easily described by passing to Fourier space and inserting the 3-cocycle phase factors
\begin{eqnarray}
\Phi_{k,k',k''}\=\e^{-\frac{\ii\hbar^2}{6}\, \rho \, \mbf k_{\mbf p}\cdot (\mbf k_{\mbf p}'\times \mbf k_{\mbf p}'')} \ , 
\end{eqnarray}
depending only on the Fourier wavenumbers $\mbf k_{\mbf p}$ dual to momentum coordinates $\mbf p$, into the Fourier transformations of triple star products. The associators satisfy `pentagon relations' which are captured by the commuting diagram
\begin{eqnarray}
\xymatrix{
 & ( {A} \star {B} )\star(
 {C} \star {D} ) \ar[ddl]_{\Phi_{{A} \star {B},
     {C}, {D}\ }}
 & \\
 & & \\
\big(( {A} \star {B} )\star
{C} \big)\star {D}
   & & {A} \star \big( {B} \star
( {C} \star {D} ) \big) \ar[uul]_{\Phi_{{A},{B},
     {C}\star {D}}} \ar[dd]^{1 \otimes
   \Phi_{{B},
     {C}, {D}}}\\
 & & \\
\big( {A} \star ( {B} \star
{C}) \big)\star {D} \ 
\ar[uu]^{\Phi_{{A},{B}, 
     {C}} \otimes 1} & & \ {A} \star
   \big((  {B} \star
{C}) \star {D}  \big) \ar[ll]^{\Phi_{{A},{B} \star 
     {C}, {D}}}
}
\end{eqnarray}
and Mac~Lane's coherence theorem asserts that this uniquely defines the insertion of suitable associator factors into higher order iterated star products. This principle enables the extension of our considerations here to the construction of some nonassociative quantum field theories~\cite{Mylonas:2014kua,Barnes:2016cjm}.

\subsection{Momentum space quantization}

Let us now examine some of the surprising physical consequences of
nonassociativity. One of the standard results in ordinary quantum
mechanics is the statement that a pair of observables which do not
commute with each other cannot be simultaneously diagonalised. Here we
find a higher version of this statement~\cite{Mylonas:2013jha}: Nonassociating observables
cannot have common eigen-states. In particular, this applies to the
diagonalisation of the basic coordinate operators {$x^I\star
  W_\psi= \lambda^I\, W_\psi$}, with {$x^I\in(\mbf x,\mbf p)$}. From
\eqref{eq:p123star}, we find that the components of momentum $\mbf p$
cannot be simultaneously measured, which implies a coarse-graining of
the momentum space with a uniform monopole background.

We can quantify this quantisation more precisely by defining oriented
area uncertainty operators
\begin{eqnarray} 
{\cal A}^{IJ} \= \mathrm{Im}\big([\widetilde x^I, \widetilde x^J]_\star\big) \= -\ii\big({\widetilde x^I} \star {\widetilde x^J} - {\widetilde x^J} \star {\widetilde x^I}\big)
\end{eqnarray}
which mimick the classical formula for the area of a triangle in terms of the vector
product between coordinate vectors in directions $I$ and
$J$, where we defined the shifted coordinates {$\widetilde x^I := x^I - \langle x^I
  \rangle_\psi$} appropriate to the computation of quantum
uncertainties. Similarly, we define oriented volume uncertainty operators
\begin{eqnarray}
{\cal V}^{IJK} \= \frac13 \, \mathrm{Re} \big( \widetilde
x^I\star [\widetilde x^J,\widetilde x^K]_{\star}+\widetilde
x^K\star [\widetilde x^I,\widetilde x^J]_{\star}+\widetilde
x^J\star [\widetilde x^K,\widetilde x^I]_{\star}\big) 
\end{eqnarray}
which mimick the classical formula for volume of a tetrahedron in terms of the triple
scalar product of coordinate vectors in directions $I$, $J$ and $K$. 

One can straightforwardly compute the expectation values of these
operators to obtain the non-vanishing minimal areas~\cite{Mylonas:2013jha}
\begin{eqnarray}
\langle {\cal A}^{x^i,p_j} \rangle_\psi \= \hbar \, \delta^i{}_j \quad
  , \quad \langle {\cal A}^{p_i,p_j} \rangle_\psi \= \frac\hbar{3} \,  \rho\,
  \varepsilon_{ijk} \,\langle x^k\rangle_\psi \ .
\end{eqnarray}
The first equation simply displays the standard Planck cells in phase
space of area $\hbar$ arising from the Heisenberg uncertainty
principle of quantum mechanics. The second equation is
new, and is due to the uncertainty between momentum measurements
proportional to the position measurement in the direction
perpendicular to the plane of the momenta, which itself is subject to the standard Planck cell uncertainty. Similarly, one obtains the
non-vanishing minimal volume~\cite{Mylonas:2013jha}
\begin{eqnarray}
\langle {\cal V}^{p_1,p_2,p_3} \rangle_\psi \= \frac12 \, \hbar^2\,  \rho
\end{eqnarray}
{which clearly indicates a quantized momentum space with a quantum of minimal volume $\frac12 \,
  \hbar^2\,  \rho$}. This forbids, in particular, configurations with
definite localised momentum in the monopole background. 

\subsection{Testable quantum effects}

Let us now take a brief interlude to consider the question
of whether it is possible to see experimental signatures of
nonassociativity in quantum
mechanics. Following~\cite{Bojowald:2015cha}, we add a confining force
with harmonic oscillator potential
\begin{eqnarray}
V(\mbf x) \= \frac12 \,m\,
  \omega^2\, \mbf x^2 \ ,
  \end{eqnarray}
and compute the effective potential due to
quantum corrections by 
\begin{eqnarray}
V_{\rm eff} \ := \ \langle{\sf H}\rangle_\psi\big|_{\langle
  p_i\rangle_\psi=0} \ .
\end{eqnarray}
In the present case, the expectation values of the Hamiltonian \eqref{eq:Hamgeneral} can be
computed by using the $\rho$-twisted Bopp shifts in
\eqref{eq:monopolestar} to write its action on a phase space wavefunction
$\psi(\mbf x,\mbf p)$ as
\begin{eqnarray}
{\sf H}(\mbf x,\mbf p)\star \psi(\mbf x,\mbf p)\= \hat{\sf H}\,\psi(\mbf
  x,\mbf p) \ ,
\end{eqnarray}
where $\hat{\sf H}$ is the second order differential
operator
\begin{eqnarray}
\hat{\sf H} &=& \frac1{2m}\, \Big(\mbf p^2+\ii\hbar\, \mbf
p\cdot\mbf\nabla_{\mbf x}-\frac{\hbar^2}{4} \, \mbf\nabla_{\mbf x}^2+
                \frac{\ii\hbar}3\,\rho \, \mbf p\cdot(\mbf
                x\times\mbf\nabla_{\mbf p})- \frac{\hbar^2}6 \,\rho\, \mbf
x\cdot(\mbf\nabla_{\mbf x}\times\mbf\nabla_{\mbf p}) \nonumber \\ &&
                                                                     \quad \qquad -\, \frac{\hbar^2}{36} \, \rho^2\, \big( (\mbf
x\cdot\mbf\nabla_{\mbf p})^2-\mbf x^2\,\mbf \nabla_{\mbf p}^2\big)
                                                                     \Big) + \frac12\, \omega^2\, \Big(\, \mbf x^2-\ii\hbar\,
\mbf x\cdot \mbf\nabla_{\mbf p}-\frac{\hbar^2}4\,
\mbf\nabla_{\mbf p}^2 \Big) \ . 
\label{eq:hatH}\end{eqnarray}

One can then express the quantum
corrections in terms of the fluctuations {$(\Delta x^I)^2 := \big\langle
  \widetilde{x}{}^I\star\widetilde{x}{}^I\big\rangle_\psi$} to
get
\begin{eqnarray}
V_{\rm eff} \= V\big(\langle\mbf x\rangle_\psi\big)+
  \frac1{2m}\, (\Delta\mbf
  p)^2+\frac12\,m\,\omega^2\, (\Delta\mbf x)^2 \ .
\end{eqnarray}
Following the standard prescription known from the associative case, the uncertainty moments can be obtained by solving Ehrenfest-type
equations of motion derived from the time evolution equations
\begin{eqnarray}
\frac{\dd \langle A\rangle_\psi}{\dd t} \= \frac{ \langle[A,{\sf
  H}]_{\star}\rangle_\psi}{\ii\hbar} 
\end{eqnarray}
in an adiabatic approximation, and in the semi-classical limit one obtains~\cite{Bojowald:2015cha} 
\begin{eqnarray}
V_{\rm eff}\= V\big(\langle\mbf x\rangle_\psi\big)+
\frac1{6m}\, \hbar \,\rho\, \big|\langle\mbf
x\rangle_\psi \big| + \frac1{2}\, \hbar\, \omega \ .
\end{eqnarray}
Hence in addition to the usual zero point energy shift of the quantum
harmonic oscillator, the effective potential demonstrates that the
motion of electrons in a magnetic monopole density exhibits anharmonic deviations
  from the classical harmonic motion. 

\begin{figure}[htb]
\begin{center}
\includegraphics[width=7cm]{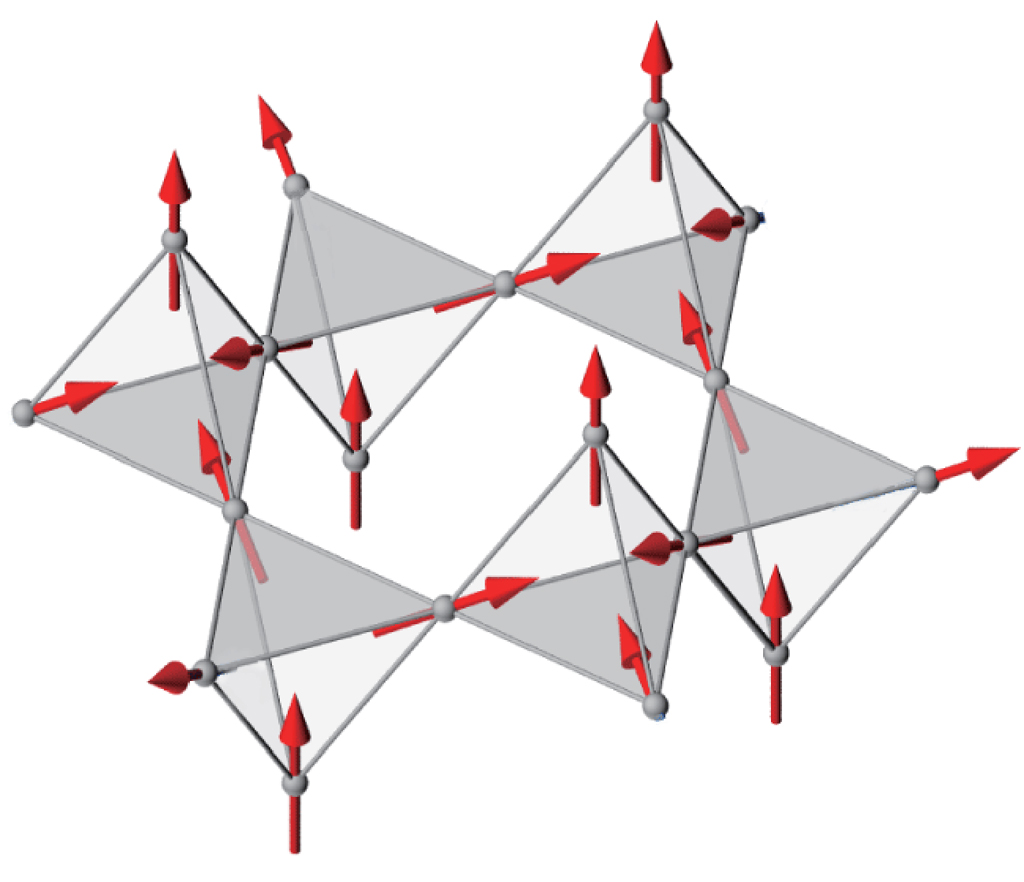}
\end{center}
\caption{\label{fig:spinice}Spin ice pyrochlore lattice with magnetic dipoles (taken from~\cite{Machida}).}
\end{figure}

This effect could have potentially observable consequences
in certain analogue systems of magnetic monopoles in condensed matter physics,
see e.g.~\cite{Castelnovo:2007qi,Morris:2010ma,Ray}. These experiments
use rare earth oxide insulators R$_2$M$_2$O$_7$ where R is a magnetic
ion (such as dysprosium or holmium) and M is a non-magnetic ion (such
as titanium). The rare earth atoms R sit at the vertices of two intertwining pyrochlore lattices formed by corner-sharing tetrahedra, as illustrated in Figure~\ref{fig:spinice}. Electron spin provides a magnetic dipole at each vertex  atom, which is shared by two regular tetrahedra, giving the lattice the geometry of quantum spin ice. The ground state contains two inward and two outward pointing dipoles towards the center of each tetrahedron, and hence has no magnetic charge. Flipping a magnetic dipole at one vertex produces a local topological excitation in which one of the two tetrahedra meeting there has an extra dipole pointing inwards while the other has an extra dipole pointing outwards, giving one tetrahedron three north poles and one south pole, and its neighbouring tetrahedron three south poles and one north pole.

Iterating this process thus gives an interpretation of flipped magnetic dipoles in terms of tetrahedra containing free and unconfined magnetic monopoles, leading as before to translation group 3-cocycles which are given by the magnetic charge inside a tetrahedron. Scattering neutrons off the material reorganises the magnetic dipoles into a 
``spin-spaghetti'' network of Dirac strings connecting pairs of
monopoles, through which magnetic flux flows. These can be
observed with applied magnetic fields at low temperature through
interference effects by their interaction with the neutrons, which
themselves carry a magnetic dipole moment. In such
experimental scenarios, magnetic monopoles therefore exist as emergent
states of matter. While these configurations do not involve a constant monopole density as in our theoretical framework, it is conceivable that the dynamics of electrons moving in a spin ice pyrochlore lattice could be approximated by electrons moving in a uniform magnetic charge distribution on scales larger than the lattice spacing, whereas the effects of Dirac strings may survive on averaging to a continuous distribution.

\newsection{Jordanian quantum mechanics and the octonions\label{sec:Jordan}}

We shall now return to some general algebraic considerations and compare the magnetic
monopole algebra with the nonassociative algebras that arose in the
early days of quantum mechanics. This will lead us into discussing the
algebra of 
octonions, which we shall show is related to the monopole
algebra via a contraction. By suitably interpreting the former algebra in this
context, we will thereby arrive at our nonassociative deformation of
three-dimensional quantum gravity.

\subsection{Noncommutative Jordan algebras}

The idea of nonassociativity in quantum mechanics is not new, and
traces back to the beginning days of quantum theory through the work
of Jordan~\cite{Jordan}, who attempted to assemble quantum observables into an
algebraic framework. The basic observation of Jordan is the
following. If $A$ and $B$ are Hermitian operators, then their product $A\, B$ is
not a Hermitian operator unless they commute, i.e. $A\, B$ is not observable unless $A$ and $B$ can be simultaneously measured; thus Hermitian operators do not close to an algebra under the
usual operator product. And neither is their commutator
$[A,B]$ an observable. However, the symmetrised product
\begin{eqnarray}
A\circ B \ := \ 
  \frac12 \,(A\,B+B\, A)
\label{eq:Jordanspecial}\end{eqnarray}
 is a Hermitian operator. The product $\circ$
is evidently commutative, 
\begin{eqnarray}
A\circ B =B\circ A \ , 
\end{eqnarray}
but it is
not associative. However, it still satisfies the `Jordan identity'
\begin{eqnarray}
(A^2\circ B)\circ A \= A^2\circ(B\circ A) \ .
\label{eq:A2BA}\end{eqnarray}
A commutative nonassociative algebra over $\bbr$ whose product $\circ$
satisfies \eqref{eq:A2BA} for all elements $A$ and $B$ is called a (linear)
{Jordan algebra}. If the algebra can be embedded in an associative algebra such that its nonassociative product $\circ$ is given as in \eqref{eq:Jordanspecial} then the Jordan algebra is
said to be ``special''. Jordan's hope was to find a system which was not derived from an associative algebra in this way, but behaved as one, in order to dispell of the insufficiencies that refer to the underlying unobservable operator algebras.

The theory of Jordan algebras was subsequently developed into a solid
mathematical component of algebra, see e.g.~\cite{Albert,Schafer},
wherein it was shown that it was sufficient to replace \eqref{eq:A2BA}
with the weaker condition that the algebra be ``alternative'':
\begin{eqnarray}
(A\circ B)\circ A\=A\circ (B\circ A) \ ,
\end{eqnarray}
and this can be used to define the notion of a {noncommutative Jordan
  algebra}. Jordan's hope of reformulating quantum mechanics in this
algebraic framework was dashed by the theorem, proven originally for
finite-dimensional algebras in~\cite{JvNW} and extended to the
infinite-dimensional case in~\cite{Zelmanov}, which essentially ruled
out the existence of non-trivial Jordan algebras: The only non-special
Jordan algebra (up to isomorphism) is the algebra of $3\times3$
Hermitian matrices over the division algebra of octonions
{$\mathbb{O}$}, which defines the 27-dimensional ``exceptional'' Jordan algebra; the octonion algebra will be described in more detail
below where it will play a prominent role. This somewhat exotic ``octonionic quantum mechanics'' satisfies the von~Neumann axioms of quantum
theory~\cite{Gunaydin:1978jq}, despite the absence of a
Hilbert space formulation. For reviews and further details, see e.g.~\cite{McCrimmon,Baez:2011hu}.

A natural question at this stage, in view of our previous version
of nonassociative quantum mechanics, is the following: Is the magnetic
monopole algebra a noncommutative Jordan algebra? This question has
been answered in the negative, originally by~\cite{Held}, and subsequently
via a more systematic treatment in~\cite{Bojowald:2016lnl} (see
also~\cite{Kupriyanov:2017oob}). The simplest counterexample already
appears at quadratic order in the phase space coordinates~\cite{Held}:
Whereas one generically has the alternativity relations
\begin{eqnarray}
(x^I\star x^K)\star x^I\=x^I\star(x^K\star x^I)
\quad , \quad \big((x^I)^{\star2}\star x^K \big)\star x^I
  \=(x^I)^{\star2}\star(x^K\star x^I) \ ,
\end{eqnarray}
the nonvanishing associator
\begin{eqnarray}
(\mbf p^2\star\mbf p^2)\star\mbf p^2 - \mbf p^2\star(\mbf p^2\star\mbf
  p^2)\= -\frac{2\ii}9 \, \hbar^2\, \rho^2\,\mbf x\cdot \mbf p \ \neq \ 0
\end{eqnarray}
demonstrates that the monopole star product does not define an
alternative algebra.

This violation of alternativity is of course related to the volume
quantization of momentum space that we observed earlier, and it brings
into question if all is not lost in this theory; for example,
this result would appear to rule out the existence of free stationary
states. However, one only needs to ensure that {$\langle\mbf x\cdot
  \mbf p\rangle_\psi=0$} in the preparation of states for
measurement. An example of such a state in the case of a free
particle, $\rho=V=0$, can be constructed from phase space
wavefunctions $\psi(\mbf x,\mbf p)$ that solve the Schr\"odinger equation
\begin{eqnarray}
\frac1{2m} \, \mbf p^2\star \psi(\mbf x,\mbf p)\=\frac1{2m}\, \hat{\mbf p}^2\psi(\mbf x,\mbf p)\= E\, \psi(\mbf x,\mbf p) \ ,
\label{eq:freeHam}\end{eqnarray}
where the first order differential operators 
\begin{eqnarray}
\hat p_i\= p_i+\frac{\ii\hbar}2\,
\frac\partial{\partial x^i}
\end{eqnarray}
are mutually commuting Hermitian operators, so they
have simultaneous eigenfunctions $\psi_{\mbf k}(\mbf x,\mbf p)$ with
\begin{eqnarray}
\hat p_i \psi_{\mbf k}(\mbf x,\mbf p) \= k_i\, \psi_{\mbf k}(\mbf
x,\mbf p) 
\end{eqnarray}
and eigenvalues $k_i\in\bbr$ for $i=1,2,3$. These three equations are solved by
\begin{eqnarray}
\psi_{\mbf k}(\mbf x,\mbf p)\=\e^{-\frac{2\ii}\hbar\, (\mbf k-\mbf p)\cdot\mbf x} \
\varphi(\mbf p)
\end{eqnarray}
where $\varphi(\mbf p)$ is an arbitrary function of $\mbf p$ independent of the
position coordinates $\mbf x$. Substituting into \eqref{eq:freeHam} then
gives the expected energy eigenvalues
\begin{eqnarray}
E\=E_{\mbf k} \=\frac{\mbf k^2}{2m} \ ,
\end{eqnarray}
and this represents the quantization of the free particle in the phase
space formulation of quantum mechanics.

Equivalently, the Schr\"odinger equation may be cast as a star-genvalue equation
\begin{eqnarray}
\frac1{2m}\, \mbf p^2 \star W_\psi\=E\, W_\psi
\end{eqnarray}
for the real-valued density operator $W_\psi(\mbf x,\mbf p)$ (see e.g.~\cite{Zachos:2001ux}). This collapses the equation to a pair of partial differential equations, its real and imaginary parts. The imaginary part
\begin{eqnarray}
\mbf p\cdot\mbf\nabla_{\mbf x}W_\psi(\mbf x,\mbf p)\=0
\end{eqnarray}
restricts $W_\psi(\mbf x,\mbf p)=W_\psi(\mbf p)$ to be independent of
$\mbf x$. The real part
\begin{eqnarray}
\Big(\mbf p^2-\frac{\hbar^2}4\, \mbf\nabla_{\mbf x}^2-2\,m\,
E \Big)W_\psi(\mbf p) \= 0
\end{eqnarray}
is satisfied for arbitrary real functions $W_\psi(\mbf p)$ of momentum with the energy eigenvalues
\begin{eqnarray}
E\=\frac{\mbf p^2}{2m} \ .
\end{eqnarray}

When a non-vanishing 
background monopole charge $\rho$ is turned on, we can consider again separately the real and imaginary parts of the
corresponding star-genvalue equation \eqref{eq:stargenvalue}. From \eqref{eq:hatH} with $\omega=0$ the imaginary equation can be written as
\begin{eqnarray}
\frac1{2m}\, \mbf p\cdot \hat{\mbf D}\,W_{\psi}(\mbf x,\mbf p)\=0
\end{eqnarray}
where
\begin{eqnarray}
\hat{\mbf D}\= \mbf\nabla_{\mbf x}- \frac\rho{3} \, \mbf
x\times\mbf\nabla_{\mbf p} \ .
\end{eqnarray}
The solutions $W_\psi(\mbf x,\mbf p)$ of this equation are the union of the integral
curves of the vector field $(\frac1m\, \mbf p,\frac\rho{3m} \,\mbf p\times \mbf x)$ on phase space. The
corresponding characteristic equations are
\begin{eqnarray}
m\, \dot{\mbf x}\= \mbf p \quad , \quad m\, \dot{\mbf p}\= \frac{\rho}3 \, \mbf p\times \mbf x \ ,
\label{eq:characteristics}\end{eqnarray}
which are simply the classical equations of motion in the monopole
background. One now needs to find from these flow equations a triple of classical integrals of motion $\mbf I=(I_1,I_2,I_3)$, such that the general solution is $W_\psi(\mbf x,\mbf p)=W_\psi(\mbf I)$, but such a complete set does not seem to exist~\cite{Bakas:2013jwa} (for the free particle with $\rho=0$ we took $\mbf I=\mbf p$).

\subsection{Octonions and magnetic monopoles}

\begin{figure}[htb]
\begin{center}
\includegraphics[width=7cm]{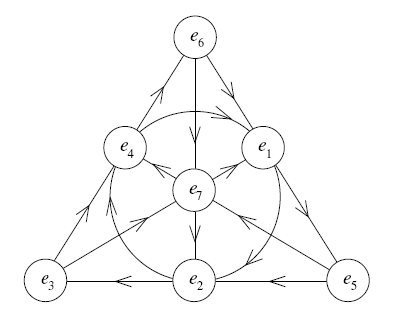}
\end{center}
\caption{\label{fig:Fano}Fano plane pneumonic for multiplication of imaginary unit octonions $e_1,\dots,e_7$.}
\end{figure}

Let us now raise the question of whether there is {\it any} relation
of the magnetic monopole algebra to Jordanian quantum mechanics. This
question was answered affirmatively by~\cite{Gunaydin:2016axc}, but to
describe that result we first need to take a slight algebraic
detour. A foundational result in algebra states that there are only
four normed division algebras over the field of real numbers: The real
numbers $\bbr$ themselves, the complex numbers $\bbc$, the quaternions
  $\mathbb{H}$, and the octionions $\mathbb{O}$. They fit
into a hierarchy $\bbr\subset\bbc\subset\mathbb{H}\subset\mathbb{O}$
where the first two
  algebras are commutative and associative, the third is
  noncommutative but associative, while the last one is both
  noncommutative and nonassociative but is alternative. An element of
  the octonion algebra $\mathbb{O}$ is a linear combination of generators
\begin{eqnarray}
a_0\, 1+a_1\, e_1+a_2\,
    e_2 +\cdots + a_7\, e_7
\end{eqnarray}
where $a_i\in\bbr$, the element $1$ is central and a unit for
the algebra, and the seven imaginary unit octonions $e_i$, with
$e_i^2=-1$ and $e_i\,e_j=-e_j\,e_i$ for $i\neq j$, have a multiplication rule which can be represented
diagrammatically through a pneumonic of the Fano plane, a finite
projective plane with seven points and seven lines, as illustrated in Figure~\ref{fig:Fano}. Each line contains three points, and each of these triples has a cyclic ordering such that if $e_i$, $e_j$ and $e_k$ are cyclically ordered in this way then
\begin{eqnarray}
e_i\,e_j\=e_k\=-e_j\,e_i \ . 
\end{eqnarray}
This defines an alternative multiplication which makes $\mathbb{O}$ into a 
finite-dimensional noncommutative Jordan algebra; see
e.g.~\cite{Baez:2011hu} for a pedagogical introduction.

For our purposes it is most convenient to relabel the imaginary units as $e_i,f_i,e_7$ with
$i=1,2,3$ and $f_1,f_2,f_3=e_4,e_5,e_6$, and to represent the multiplication
rule by
\begin{eqnarray}
{e_i\,e_j}&{=}&{-\delta_{ij} + \varepsilon_{ijk}\, e_k} \qquad , \qquad
                {e_7\,e_i} \ \ = \ \ {
             f_i} \ , \nonumber \\[4pt]
{} {f_i\,f_j}&{=}&{-\delta_{ij}- \varepsilon_{ijk}\, e_k \qquad , \qquad
                   e_7\,f_i} \ \ = \ \ 
             { e_i} \ , \nonumber\\[4pt]
{} {e_i\,f_j}&{=}&{\delta_{ij}\, e_7- \varepsilon_{ijk}\, f_k}  \ . 
\label{eq:octonionmult}\end{eqnarray}
The first relation shows that $e_i$ for $i=1,2,3$ generate an associative
quaternion subalgebra $\mathbb{H}\subset\mathbb{O}$. The non-vanishing
Jacobiators are given by
\begin{eqnarray}
[e_i,e_j,f_k]&=&4\, (\varepsilon_{ijk}\,
                                e_7+ \delta_{kj}\,
                          f_i- \delta_{ki}\, f_j ) \ ,
                                 \nonumber \\[4pt]
[e_i,f_j,f_k]&=&-4\, ( \delta_{ij}\, e_k-\delta_{ik}\, e_j)
                           \ ,\nonumber\\[4pt]
[f_i,f_j,f_k]&=&-4\, \varepsilon_{ijk}\, e_7 \ ,\nonumber\\[4pt]
[e_i,e_j,e_7]&=&-4\, \varepsilon_{ijk}\, f_k \ ,\nonumber\\[4pt]
[e_i,f_j,e_7]&=&-4\, \varepsilon_{ijk}\, e_k \ ,\nonumber\\[4pt]
[f_i,f_j,e_7]&=&4\, \varepsilon_{ijk}\, f_k \ . 
\label{mp2}\end{eqnarray}

Following \cite{Gunaydin:2016axc}, let us now rescale the octonionic
units and set
\begin{eqnarray}
x^i\=-\frac{\ii\hbar}2 \, \lambda\, e_i \quad , \quad
  p_i\=\frac{\ii\hbar}{2 \sqrt3} \, \sqrt{\lambda\,\rho}\ f_i \quad ,
  \quad I\=\frac{\ii\hbar}{2 \sqrt3} \, \sqrt{\lambda^3\, \rho} \ e_7
\label{eq:contrsub}\end{eqnarray}
for a parameter $\lambda\in\bbr$, and take the contraction limit
$\lambda \rightarrow 0$ of the octonionic commutation relations
following from \eqref{eq:octonionmult}. The various commutators then
contract in the following way:
\begin{eqnarray}
[e_i,e_j]\=2\, \varepsilon_{ijk}\, e_k \quad
  &\stackrel{\lambda\to0}{\Longrightarrow}& \quad [x^i,x^j]\=0 \ ,
  \nonumber \\[4pt]
[f_i,f_j]\=-2\, \varepsilon_{ijk}\, e_k \quad
  &\stackrel{\lambda\to0}{\Longrightarrow}& \quad
  [p_i,p_j]\=\frac{\ii\hbar}3 \,\rho\,\varepsilon_{ijk}\, x^k \ , \nonumber \\[4pt]
[e_i,f_j]\=2\, (\delta_{ij}\, e_7- \varepsilon_{ijk}\,
    f_k) \quad &\stackrel{\lambda\to0}{\Longrightarrow}& \quad
  [x^i,p_j]\=\ii\hbar\, \delta^i{}_j\, I \ , \nonumber \\[4pt]
[e_7,e_i]\=2\,
             f_i \ , \ [e_7,f_i] \=
             -2\, e_i \quad &\stackrel{\lambda\to0}{\Longrightarrow}& \quad
           [x^i,I]\= 0 \= [p_i,I] \ .
\end{eqnarray}
Moreover, only one Jacobiator from \eqref{mp2} contracts non-trivially
and is given by
\begin{eqnarray}
[f_1,f_2,f_3]\=-4\, e_7 \quad &\stackrel{\lambda\to0}{\Longrightarrow}&
  \quad [p_1,p_2,p_3]\=
    \hbar^2\,\rho\, I \ .
\end{eqnarray}
This demonstrates a surprising relationship between magnetic monopoles
and the octonions~\cite{Gunaydin:2016axc}: The contracted octonion algebra is the magnetic
monopole algebra given by \eqref{eq:Lorentzcommrels} and
\eqref{eq:mom3bracket}, with central element $I$. This result was extended
by~\cite{Kupriyanov:2017oob} to a complete quantization of algebras of
functions by an explicit construction of an exact (though somewhat
complicated) star product which quantizes the octonions, and contracts
non-trivially to the monopole star product \eqref{eq:monopolestar}.

\subsection{Nonassociative quantum gravity}

By relabelling $I=p_4$, the uncontracted octonion algebra reads as
\begin{eqnarray}
{[x^i,x^j]} &{=} &{\ii\hbar\,\lambda\,
                                 \varepsilon^{ijk}\, x^k} \ ,
                   \nonumber \\[4pt]
{[p_i,p_j]} &{=}&{\frac{\ii\hbar}3 \,\rho \, \varepsilon_{ijk}\, x^k  \qquad , \qquad
                        [p_4,p_i] \ \ = \ \ -\frac{\ii\hbar}3 \,\lambda\, \rho\,
                  x_i} \ , \nonumber \\[4pt]
{} {[x_i,p_j]} &{=}& {\ii\hbar\, \delta_{ij}\, p_4-\ii\hbar\, \lambda\,
                        \varepsilon_{ijk}\, p_k \qquad ,
                        \qquad [x^i,p_4] \ \ = \ \ -\ii\hbar\,
                     \lambda^2\,x^i } \ .
\label{eq:uncontroctonion}\end{eqnarray}
This algebra describes a seven-dimensional nonassociative phase space
with an ``extra'' momentum mode 
  ${p_4}$; the non-trivial Jacobiators can be read off by substituting
  \eqref{eq:contrsub} into \eqref{mp2}. Let us now elucidate the physical meaning of this phase
  space algebra, and in particular the contraction
  parameter ${\lambda}$, following~\cite{Lust:2017bgx}.
  
Looking at the quaternionic (or $SU(2)$) subalgebra generated by the position coordinates
$x^i$ in \eqref{eq:uncontroctonion}, setting ${\rho=0}$ reveals a
  noncommutative but associative deformation of spacetime with 
  ${[p_i,p_j]=0}$. The combination ${\lambda^2\,\mbf
  p^2+p_4^2}$ is straightforwardly calculated to be a central element
of this algebra. Setting it to $1$, and then restricting the momenta
to the upper hemisphere of this sphere to eliminate $p_4$, the
remaining non-vanishing commutation relations from
\eqref{eq:uncontroctonion} become
\begin{eqnarray}
{[x^i,x^j] \= \ii\hbar\, \lambda\, \varepsilon^{ijk}\,
x^k \quad , \quad [x_i,p_j] \=
\ii\hbar\,
\sqrt{1-\lambda^2\, \mbf p^2} \ \delta_{ij} - \ii\hbar\,
\lambda\, \varepsilon_{ijk}\, p_k} \ .
\end{eqnarray}
We therefore recover in this way the deformed phase space commutation
relations \eqref{eq:3DQGrels} of three-dimensional quantum gravity provided we identify
\begin{eqnarray}
\lambda\=\frac{\ell_{\rm P}}{\hbar} \ .
\label{eq:lambdaPlanck}\end{eqnarray}

This observation suggests a novel nonassociative deformation of
three-dimensional quantum gravity, wherein the uncontracted octonion
algebra \eqref{eq:uncontroctonion} appears to be related to monopoles in
  the spacetime of three-dimensional quantum gravity, with the
  contraction parameter identified with the Planck length through \eqref{eq:lambdaPlanck}. 
The meaning of the contraction limit $\lambda\to0$ is then clear: It
corresponds to turning off quantum gravitational effects, leaving the
nonassociative quantum mechanics of electrons in the
background of a constant magnetic charge density. 

\newsection{Epilogue: Magnetic monopoles in M-theory\label{sec:KKmonopole}}

The interpretation of the nonassociative phase space algebra
\eqref{eq:uncontroctonion} as describing the
quantum mechanics of electrons propagating simultaneously in
magnetic charge and gravitational backgrounds is very tantalising in
light of recent suggestions from non-geometric string theory, which
anticipate a nonassociative theory of gravity to govern the low-energy
effective dynamics of closed strings in certain locally non-geometric
backgrounds~\cite{Blumenhagen:2010hj,Blumenhagen:2013zpa,Aschieri:2015roa,Barnes:2015uxa,Barnes:2016cjm,Blumenhagen2016}. However,
the appearence of the three-sphere in momentum space, which is a
characteristic feature of three-dimensional quantum gravity, is put in
here by hand as a restriction of a four-dimensional momentum
space. This suggests that a deeper structure is inherent in the physics
governed by the phase space relations \eqref{eq:uncontroctonion}, and
it remains to find a concrete physical realisation.

One possible such scenario was proposed by~\cite{Lust:2017bgx} in the framework
of M-theory, which by definition is
a quantum theory of gravity in 11 dimensions. The starting point is to embed magnetic monopoles into IIA string theory as D6-branes, where the corresponding electron probes are D0-branes. The D6-brane lifts to M-theory as a Kaluza-Klein monopole described by the 11-dimensional metric
\begin{eqnarray}
{\dd s_{11}^2\=\dd s_7^2+H\,\dd\mbf x\cdot\dd\mbf x+H^{-1}\,\big(\dd
x^4+\mbf A\cdot\dd\mbf x\big)^2} \ ,
\label{eq:KKmetric}\end{eqnarray}
where $x^4\in S^1$ is the M-theory direction and the harmonic function $H(\mbf x)$ is defined by
\begin{eqnarray}
{\mbf\nabla\times\mbf A\=\mbf\nabla H \quad , \quad
\mbf\nabla^2H\=\rho } \ .
\end{eqnarray}
The corresponding electron probes lifting D0-branes are M-waves
  along $S^1$, which are graviton momentum modes with
\begin{eqnarray}
p_4\=\frac{\hbar\,e}{R_{11}} \ .
\label{eq:gravitonp4}\end{eqnarray}
Because the probes are waves, they have no local position along the
M-theory circle; hence the coordinate $x^4$ does not appear in the phase space algebra \eqref{eq:uncontroctonion}, which we propose to describe the M-waves in the Kaluza-Klein monopole background.

The relevant parameters in IIA string theory are the string length $\ell_s$ and the string coupling $g_s$, while those of M-theory are the 11-dimensional Planck length $\ell_{\rm P}$ and the radius $R_{11}$ of the M-theory circle. They are related through
\begin{eqnarray}
\ell_s^2 \=\frac{\ell_{\rm
      P}^3}{R_{11}} \quad , \quad g_s\=\bigg(\frac{R_{11}}{\ell_{\rm
        P}}\bigg)^{3/2} \ .
\end{eqnarray}
The reduction of M-theory to IIA string theory is defined by the limit ${g_s,R_{11}
\rightarrow 0}$  with ${\ell_s}$ finite. With the identification \eqref{eq:lambdaPlanck}, this implies the contraction limit ${\lambda\sim\ell_{\rm P}\sim
  R_{11}^{1/3}\rightarrow 0}$ reducing the M-wave algebra \eqref{eq:uncontroctonion} to the magnetic monopole algebra \eqref{eq:Lorentzcommrels} of the D0-branes. In this limit the graviton momentum mode \eqref{eq:gravitonp4} is frozen and hence disappears from the phase space, while the Kaluza-Klein quantum number $e$ becomes the electric charge of the D0-branes. 

For a single Dirac monopole, wherein $\rho$ is a delta-function distribution, the four-dimensional part of \eqref{eq:KKmetric} is the metric of the hyper-K\"ahler Taub-NUT space with local coordinates $(\mbf x,x^4)\in\bbr^3\times S^1$. For the uniform distribution of magnetic charge that we are interested in here, $\rho$ can be written as an integral over infinitely-many densely distributed Dirac
  monopoles. In this case, there is no local expression for the vector potential ${\mbf A}$ and hence no local expression for the metric \eqref{eq:KKmetric}; the resulting solution can thereby be thought of as a `non-geometric Kaluza-Klein monopole'. Similary to what occured before, the emergence of local non-geometry here arises because the local $S^1$-fibration over $\bbr^3$ described by the Taub-NUT space is replaced by an $S^1$-gerbe over ${\bbr^3}$. For further details, see~\cite{Lust:2017bgx}.

\ack

The author thanks Peter Schupp and Vladislav Kupriyanov for helpful
discussions, and Cestmir Burdik for the invitation to present this
material at ISQS25. This work was supported by the COST Action MP1405 QSPACE, funded by the
European Cooperation in Science and Technology (COST), and by the
Consolidated Grant ST/L000334/1 from the UK Science and Technology
Facilities Council (STFC).

\section*{References}


\begin{thebibliography}{10}
\expandafter\ifx\csname url\endcsname\relax
  \def\url#1{{\tt #1}}\fi
\expandafter\ifx\csname urlprefix\endcsname\relax\def\urlprefix{URL }\fi
\providecommand{\eprint}[2][]{\url{#2}}

\bibitem{tHooft:1996ziz}
't~Hooft G 1996 {\em Class. Quant. Grav.\/} {\bf 13} 1023--1040
  (\textit{Preprint} \eprint{gr-qc/9601014})

\bibitem{Matschull:1997du}
Matschull H~J and Welling M 1998 {\em Class. Quant. Grav.\/} {\bf 15}
  2981--3030 (\textit{Preprint} \eprint{gr-qc/9708054})

\bibitem{Freidel:2005me}
Freidel L and Livine E~R 2006 {\em Phys. Rev. Lett.\/} {\bf 96} 221301
  (\textit{Preprint} \eprint{hep-th/0512113})

\bibitem{Blumenhagen:2010hj}
Blumenhagen R and Plauschinn E 2011 {\em J. Phys. A\/} {\bf 44} 015401
  (\textit{Preprint} \eprint{1010.1263})

\bibitem{Lust:2010iy}
L{\"u}st D 2010 {\em J. High Energy Phys.\/} {\bf 12} 084 (\textit{Preprint}
  \eprint{1010.1361})

\bibitem{Blumenhagen:2011ph}
Blumenhagen R, Deser A, L{\"u}st D, Plauschinn E and Rennecke F 2011 {\em J.
  Phys. A\/} {\bf 44} 385401 (\textit{Preprint} \eprint{1106.0316})

\bibitem{Mylonas:2012pg}
Mylonas D, Schupp P and Szabo R~J 2012 {\em J. High Energy Phys.\/} {\bf 09}
  012 (\textit{Preprint} \eprint{1207.0926})

\bibitem{Bakas:2013jwa}
Bakas I and L{\"u}st D 2014 {\em J. High Energy Phys.\/} {\bf 01} 171
  (\textit{Preprint} \eprint{1309.3172})

\bibitem{Blumenhagen:2013zpa}
Blumenhagen R, Fuchs M, Hassler F, L{\"u}st D and Sun R 2014 {\em J. High
  Energy Phys.\/} {\bf 04} 141 (\textit{Preprint} \eprint{1312.0719})

\bibitem{Aschieri:2015roa}
Aschieri P and Szabo R~J 2015 {\em J. Phys. Conf. Ser.\/} {\bf 634} 012004
  (\textit{Preprint} \eprint{1504.03915})

\bibitem{Gunaydin:2016axc}
G{\"u}naydin M, L{\"u}st D and Malek E 2016 {\em J. High Energy Phys.\/} {\bf
  11} 027 (\textit{Preprint} \eprint{1607.06474})

\bibitem{Kupriyanov:2017oob}
Kupriyanov V~G and Szabo R~J 2017 {\em J. High Energy Phys.\/} {\bf 02} 099
  (\textit{Preprint} \eprint{1701.02574})

\bibitem{Lust:2017bgx}
L{\"u}st D, Malek E and Szabo R~J 2017 {Non-geometric Kaluza-Klein monopoles
  and magnetic duals of M-theory $R$-flux backgrounds} (\textit{Preprint}
  \eprint{1705.09639})

\bibitem{Zak:1964zz}
Zak J 1964 {\em Phys. Rev.\/} {\bf 134} A1602--A1606

\bibitem{Dirac}
Dirac P~A~M 1948 {\em Phys. Rev.\/} {\bf 74} 817--830

\bibitem{Jackiw:1984rd}
Jackiw R 1985 {\em Phys. Rev. Lett.\/} {\bf 54} 159--162

\bibitem{WuYang}
Wu T~T and Yang C~N 1976 {\em Nucl. Phys. B\/} {\bf 107} 365--380

\bibitem{Barnes:2014ksa}
Barnes G~E, Schenkel A and Szabo R~J 2015 {\em J. Geom. Phys.\/} {\bf 89}
  111--152 (\textit{Preprint} \eprint{1409.6331})

\bibitem{Bunk:2016rta}
Bunk S, Saemann C and Szabo R~J 2016 {The 2-Hilbert space of a prequantum
  bundle gerbe} (\textit{Preprint} \eprint{1608.08455})

\bibitem{Bunk:2016gus}
Bunk S and Szabo R~J 2017 {\em Lett. Math. Phys.\/} {\bf 107} 1877--1918
  (\textit{Preprint} \eprint{1612.01878})

\bibitem{Gunaydin:1985ur}
G{\"u}naydin M and Zumino B 1985 {\em {Old and New Problems in Fundamental
  Physics: Meeting in Honour of G.C. Wick}\/} pp 43--53

\bibitem{Gunaydin:2013nqa}
G{\"u}naydin M and Minic D 2013 {\em Fortsch. Phys.\/} {\bf 61} 873--892
  (\textit{Preprint} \eprint{1304.0410})

\bibitem{Bojowald:2014oea}
Bojowald M, Brahma S, B{\"u}y{\"u}k{\c{c}}am U and Strobl T 2015 {\em J. High
  Energy Phys.\/} {\bf 03} 093 (\textit{Preprint} \eprint{1411.3710})

\bibitem{Gronewold}
Gr{\"o}newold H 1946 {\em Physica\/} {\bf 12} 405--460

\bibitem{Moyal}
Moyal J~E 1949 {\em Proc. Cambridge Phil. Soc.\/} {\bf 45} 99--124

\bibitem{Weyl}
Weyl H 1927 {\em Z. Phys.\/} {\bf 46} 1--46

\bibitem{Wigner}
Wigner E 1932 {\em Phys. Rev.\/} {\bf 40} 749--759

\bibitem{Zachos:2001ux}
Zachos C~K 2002 {\em Int. J. Mod. Phys. A\/} {\bf 17} 297--316
  (\textit{Preprint} \eprint{hep-th/0110114})

\bibitem{Mylonas:2013jha}
Mylonas D, Schupp P and Szabo R~J 2014 {\em J. Math. Phys.\/} {\bf 55} 122301
  (\textit{Preprint} \eprint{1312.1621})

\bibitem{Kupriyanov:2015dda}
Kupriyanov V~G and Vassilevich D~V 2015 {\em J. High Energy Phys.\/} {\bf 09}
  103 (\textit{Preprint} \eprint{1506.02329})

\bibitem{Mylonas:2014kua}
Mylonas D and Szabo R~J 2014 {\em Fortsch. Phys.\/} {\bf 62} 727--732
  (\textit{Preprint} \eprint{1404.7304})

\bibitem{Barnes:2016cjm}
Barnes G~E, Schenkel A and Szabo R~J 2016 {\em Proc. Science\/} {\bf CORFU2015}
  081 (\textit{Preprint} \eprint{1601.07353})

\bibitem{Bojowald:2015cha}
Bojowald M, Brahma S and B{\"u}y{\"u}k{\c{c}}am U 2015 {\em Phys. Rev. Lett.\/}
  {\bf 115} 220402 (\textit{Preprint} \eprint{1510.07559})

\bibitem{Castelnovo:2007qi}
Castelnovo C, Moessner R and Sondhi S~L 2008 {\em Nature\/} {\bf 451} 42--45
  (\textit{Preprint} \eprint{0710.5515})

\bibitem{Morris:2010ma}
Morris D~J~P {\em et~al.\/} 2009 {\em Science\/} {\bf 326} 411--414
  (\textit{Preprint} \eprint{1011.1174})

\bibitem{Ray}
Ray M~W, Ruokokoski E, Tiurev K, M{\"o}tt{\"o}nen and Hall D~S 2015 {\em
  Science\/} {\bf 348} 544--547

\bibitem{Machida}
Machida Y, Nakatsuji S, Onoda S, Tayama T and Sakakibara T 2010 {\em Nature\/}
  {\bf 463} 210--213

\bibitem{Jordan}
Jordan P 1932 {\em Nachr. Ges. Wiss. G{\"o}ttingen\/} {\bf 1932} 569--575

\bibitem{Albert}
Albert A~A 1946 {\em Trans. Amer. Math. Soc.\/} {\bf 59} 524--555

\bibitem{Schafer}
Schafer R~D 1955 {\em Bull. Amer. Math. Soc.\/} {\bf 61} 469--484

\bibitem{JvNW}
Jordan P, von Neumann J and Wigner E 1934 {\em Ann. Math.\/} {\bf 35} 29--64

\bibitem{Zelmanov}
Zelmanov E~I 1983 {\em Sibirsk Mat. J.\/} {\bf 24} 89--104

\bibitem{Gunaydin:1978jq}
G{\"u}naydin M, Piron C and Ruegg H 1978 {\em Commun. Math. Phys.\/} {\bf 61}
  69--85

\bibitem{McCrimmon}
McCrimmon K 1978 {\em Bull. Amer. Math. Soc.\/} {\bf 84} 612--627

\bibitem{Baez:2011hu}
Baez J~C 2012 {\em Found. Phys.\/} {\bf 42} 819--855 (\textit{Preprint}
  \eprint{1101.5690})

\bibitem{Held}
Held A and Schupp P 2014 unpublished

\bibitem{Bojowald:2016lnl}
Bojowald M, Brahma S, B{\"u}y{\"u}k{\c{c}}am U and Strobl T 2017 {\em J. High
  Energy Phys.\/} {\bf 04} 028 (\textit{Preprint} \eprint{1610.08359})

\bibitem{Barnes:2015uxa}
Barnes G~E, Schenkel A and Szabo R~J 2016 {\em J. Geom. Phys.\/} {\bf 106}
  234--255 (\textit{Preprint} \eprint{1507.02792})

\bibitem{Blumenhagen2016}
Blumenhagen R and Fuchs M 2016 {\em J. High Energy Phys.\/} {\bf 07} 019
  (\textit{Preprint} \eprint{1604.03253})

\end{thebibliography}

\providecommand{\newblock}{}

\end{document}